\documentclass[%
 reprint,
superscriptaddress,
showpacs,
 amsmath,amssymb,
 aps,
floatfix,
]{revtex4-1}

\usepackage[dvipdfmx,colorlinks=true,urlcolor=blue,citecolor=blue,linkcolor=blue]{hyperref}
\usepackage[all]{hypcap}

\usepackage{dcolumn}% Align table columns on decimal point
\usepackage{bm}% bold math

\usepackage[dvipdfmx]{graphicx,color}% Include figure files

\usepackage{ulem} % sout

\begin{document}
\title{Ultrafast transient interference in pump-probe spectroscopy \\ of band and Mott insulators}
\author{Kazuya Shinjo}
%\email{shinjo@rs.tus.ac.jp}
\author{Takami Tohyama}
%\email{tohyama@rs.tus.ac.jp}
\affiliation{Department of Applied Physics, Tokyo University of Science, Tokyo 125-8585, Japan}

\date{\today}% It is always \today, today,
             

\begin{abstract}
Ultrafast pump-probe spectroscopy with high temporal and spectral resolutions provides new insight into ultrafast nonequilibrium phenomena. We propose that transient interference between pump and probe pulses is realized in pump-probe spectroscopy of band and Mott insulators, which can be observed only after recent developments of ultrafast spectroscopic techniques. A continuum structure in the excitation spectrum of band insulators is found to act as a medium for storing the spectral information of the pump pulse, and the spectrum detected by the probe pulse is interfered with by the medium, generating the transient interference in the energy domain. We also demonstrate the transient interference in the presence of electron correlations in a one-dimensional half-filled Hubbard model. Furthermore, bosons coupled to electrons additively contribute to the interference. Our finding will provide an interpretation of probe-energy-dependent oscillations recently observed in the pump-probe spectrum for a two-dimensional Mott insulator.
\end{abstract}
\maketitle

%%%%%%%%%%%%%%%%%
%
%%%%%%%%%%%%%%%%%
\section{introduction}\label{sec_1}
Ultrafast pump-probe spectroscopy is a good tool to investigate the nonequilibrium properties of a given system since a pump pulse triggers ultrafast processes and a subsequent probe pulse monitors the pump-induced dynamical processes~\cite{Mukamel1995, Diels1996, Krausz2009, Giannetti2016}.
Especially, by using femtosecond pulses, nonequilibrium dynamics of electrons can be detected since the timescale of the motion of electrons is of the order of a femtosecond.
However, increasing the resolution of optical measurements in both the time and energy domains is difficult and limited by the uncertainty principle.

Recently, ultrafast spectroscopic techniques have been advanced by using a transform-limited pulse, i.e., a pulse that has the minimum possible duration for a given spectral bandwidth, and have opened a new door to make both temporal and spectral resolutions as high as possible~\cite{Diels1996}.
These techniques can disclose new ultrafast nonequilibrium phenomena.
In fact, by applying these techniques, interference in the energy domain has been observed in atomic systems and nanometric tips~\cite{Wollenhaupt2002, Lindner2005, Kiffner2006, Milosevic2006, Kruger2018}.
This interference is applied to control the atomic storage medium for recording the information of optical pulses~\cite{Leung1982, Carlson1983, Hemmer1994, Fleischauer2002, Ohmori2009}.
However, as far as we know, there has been no such report on transient interference of pump-probe spectroscopy of band and Mott insulators both experimentally and theoretically.

In this paper, we investigate ultrafast pump-probe spectroscopy of band and Mott insulators and propose transient interference between temporary well-separated pulses in electron systems as in the case of atomic systems.
We formulate such transient interference in pump-probe spectroscopy of a two-band model.
We find that the existence of a continuum structure in the excitation spectrum is important for generating the transient interference since the continuum structure acts as a medium for storing the spectral information of the pump pulse and for creating interference between temporary well-separated pump and probe photons.
The information persists due to a memory effect, i.e., a relaxation process of electron systems.
As a result, the time-domain pump-probe spectrum depends on both probe energy $\omega$ and the central frequency of the pump and probe pulses $\Omega$ and thus oscillates with a frequency
\begin{align}
\label{probe_dep}
\omega_0 = \omega - \Omega.
\end{align}
In order to demonstrate the transient interference in the presence of electron correlation, we perform numerical calculations of the pump-probe spectrum in a one-dimensional (1D) half-filled Hubbard model.
Moreover, we find that bosons coupled to electrons in the two-band model make an additional contribution to the interference.
Based on the result, we speculate that the transient interference will be observed in Mott insulators strongly correlated to magnons.
For the observation of the proposed transient interference, high resolution of measurements of both time and energy is required in ultrafast pump-probe spectroscopy.
Recently, oscillations of electronic states above the charge-transfer gap in a two-dimensional (2D) Mott insulator Nd$_2$CuO$_4$ were observed on the reflectivity changes detected by pump-probe measurement with ultrashort pulses~\cite{Miyamoto2018}. 
The time and energy resolutions of the measurement are as high as $10 \text{fs}$ and $0.01 \text{eV}$, respectively.
By extracting the oscillatory components from the pump-probe spectrum, the oscillation component with the frequency indicated by Eq.~(\ref{probe_dep}) was found~\cite{Miyamoto2018}.
We propose that the transient interference will be one of the possible origins of the observed oscillations.

This paper is organized as follows. We introduce a two-band model, which is a minimal model to describe the interference effect by two photon pulses through an electron system, and show the pump-probe absorption spectrum in Sec.~\ref{sec_2}. In Sec.~\ref{sec_3}, we calculate the time-dependent optical conductivity at half filling just after pumping. The effect of bosons coupled to electrons on the pump-probe spectrum is discussed in Sec.~\ref{sec_4}. Finally, a summary is given in Sec.~\ref{sec_5}.

\section{Two-band model}\label{sec_2}
We first introduce a two-band model, which is the minimal model to describe the interference effect by two photon pulses through an electron system, and analytically calculate the pump-probe absorption spectrum.
With the assumption of dipole transitions, the Hamiltonian of the two-band model under the time-dependent electric field reads
\begin{align*}
\mathcal{H} =& \sum_{\bm{k}} \varepsilon_{\bm{k}} c_{\text{c}\bm{k}}^\dag c_{\text{c}\bm{k}} + \sum_{\bm{k}} \varrho_{\bm{k}} c_{\text{v}\bm{k}}^\dag c_{\text{v}\bm{k}} \nonumber\\
&-\sum_{\bm{k}} \left( d_{\text{cv}} \mathcal{E}(t) c_{\text{c}\bm{k}}^\dag c_{\text{v}\bm{k}} + d_{\text{cv}}^* \mathcal{E}(t) c_{\text{v}\bm{k}}^\dag c_{\text{c}\bm{k}} \right), \nonumber
\end{align*}
where $c_{\text{c(v)}\bm{k}}$ is an annihilation operator for fermions in the conduction (valence) band with momentum $\bm{k}$. The energies of the conduction and valence bands are $\varepsilon_{\bm{k}} = \varepsilon + \frac{\hbar ^2 \bm{k}^2}{2m_{\text{c}}}$ and $ \varrho_{\bm{k}} = \varrho + \frac{\hbar ^2 \bm{k}^2}{2m_{\text{v}}}$,
where $\varepsilon$ and $\varrho$ are the minimum and maximum of the conduction and valence bands, respectively, and $m_c$ and $m_v$ are the effective masses of electrons in the conduction and valence bands, respectively.
We introduce the interband dipole matrix element $d_{\text{cv}}$ and external electric field $\mathcal{E}(t)$. Hereafter, we set $\hbar=1$.

By taking the long-wave-length limit of the electric field, the optical Bloch equation is written as~\cite{HaugKoch}
\begin{align} \label{Eq_bloch_eq_1}
&\left( \frac{\partial}{\partial t} + i\{ \varepsilon _{\bm{k}}-\varrho_{\bm{k}}-i \gamma \} \right) p_{\text{vc}}^0(\bm{k},t) = d_{\text{cv}} \mathcal{E}(t) \{ 1-2f_c(\bm{k}) \}
\end{align}
and
\begin{align} \label{Eq_bloch_eq_2}
\left(  \frac{\partial}{\partial t} +  \Gamma \right) f_{\text{c}}(\bm{k},t) = -2\text{Im}\left[ d_{\text{cv}} \mathcal{E}(t) p_{\text{vc}}^{0*}(\bm{k},t) \right],
\end{align}
where $f_{\text{c}}(\bm{k}) = \langle c_{\text{c}\bm{k}}^\dag c_{\text{c}\bm{k}} \rangle$ and $p_{\text{vc}}^0(\bm{k}) = \langle c_{\text{v}\bm{k}}^\dag c_{\text{c}\bm{k}} \rangle$, with $\langle \cdots \rangle$ representing the expectation value.
We introduce a phenomenological damping rate $\Gamma$ for $f_{\text{c}}$ and dephasing rate $\gamma$ for $p_{\text{vc}}^0$.
We consider an electric field $\mathcal{E}(t) = \frac{1}{2}\left(\mathcal{\tilde{E}}(t)e^{-i\Omega t} + \mathcal{\tilde{E}}^*(t)e^{i\Omega t}\right)$, where $\mathcal{\tilde{E}}(t)=2\left\{ \mathcal{\tilde{E}}_{\text{p}}(t)e^{i\bm{k}_p\cdot \bm{r}} + \mathcal{\tilde{E}}_{\text{t}}(t)e^{i\bm{k}_t\cdot \bm{r}} \right\}$, and the electric field and wave vector of the pump (probe) pulse are $\mathcal{\tilde{E}}_{\text{p}}$ and $\bm{k}_{\text{p}}$ ($\mathcal{\tilde{E}}_{\text{t}}$ and $\bm{k}_{\text{t}}$), respectively.
Introducing an expansion parameter $\lambda$ through $\mathcal{E}(t) \rightarrow \lambda \mathcal{E}(t)$, we obtain $p_{\text{vc}}^0 = \lambda p_{\text{vc}}^{0(1)} + \lambda^2 p_{\text{vc}}^{0(2)} + \lambda^3 p_{\text{vc}}^{0(3)} + \cdots,\; f_{\text{c}} = \lambda f_{\text{c}}^{(1)} + \lambda^2 f_{\text{c}}^{(2)} + \lambda^3 f_{\text{c}}^{(3)} + \cdots$.
The shape of the probe pulse is represented by the delta function $\mathcal{\tilde{E}}_{\text{t}}(t) = \mathcal{\tilde{E}}_{\text{t}} \delta (t -\tau)$ $(\tau >0)$, where $\tau$ is the delay time between the pump and probe pulses.
The pump-induced absorption change is given by $\alpha = -\textrm{Im}\left[ d_{cv}^*\chi(\bm{k},\omega) \right].$
Taking $\mathcal{\tilde{E}}_{\text{p}}(t)=\mathcal{\tilde{E}}_{\text{p}}e^{-\sigma|t|}$ and with the rotating-wave approximation, the probe susceptibility is given by (see Appendix A)
\begin{align} \label{chi0}
&\chi (\bm{k},\omega)\simeq \frac{p_{\text{vc}}^{0(3)}(\bm{k},\omega)}{\mathcal{E}_{\text{t}}(\omega)}\nonumber \\
&=\frac{8d_{\text{cv}}\left|d_{\text{cv}}\right| {}^2 \left| \mathcal{\tilde{E}}_p\right| {}^2  e^{-\left(\sigma - \gamma \right)\tau} e^{i \tau  \left( - \Omega +\varepsilon _k-\varrho   _k\right)}  \Gamma \sigma }{\left(i \gamma +\omega -\varepsilon _k+\varrho _k\right)(i\Gamma +i\sigma + \omega -\Omega ) v_{\bm{k}}^+ u_{\bm{k}}^+ u_{\bm{k}}^-}+\cdots,
\end{align}
where $u_{\bm{k}}^{\pm}=i\gamma \pm i\sigma + \Omega - \varepsilon _{\bm{k}} + \varrho _{\bm{k}}$ and $v_{\bm{k}}^+ = i\gamma +i\Gamma -i\sigma + \Omega - \varepsilon _{\bm{k}}+ \varrho _{\bm{k}}$. 
In the limit $\gamma\rightarrow 0$, the pole of the energy denominator $\omega = \varepsilon_{\bm{k}} - \varrho _{\bm{k}}$ in the third term of $\chi (\bm{k},\omega)$ gives rise to an oscillatory behavior of $e^{i(\omega -\Omega)\tau}$ with decay $e^{-(\sigma-\gamma)\tau}$.
This is the oscillation component indicated by Eq.~(\ref{probe_dep}).
Since the timescale where the oscillation persists is on the order of $\gamma^{-1}$, real-time ultrafast dynamics should be observed with high accuracy~\cite{Rhodes2013}.

In order to maintain the oscillation in the two-band model, we have to select a proper set of parameters that leads to the coherence and memory effect in the energy domain.
First of all, we examine the coherence in the energy domain.
When $\sigma \gg 1/\tau$, i.e., the pulse duration is much shorter than the time delay $\tau$, we obtain $\Delta t \sim 0$, where $\Delta t$ is the uncertainty in the time domain.
Simultaneously, the energy uncertainty $\Delta E \sim \infty$, leading to low energy resolution.
As a result, the interference in the energy domain is invisible.
This corresponds to the fact that the interference pattern vanishes in Young's double-slit experiment if the path of light is measured~\cite{Englert1996, Durr1998}.
In fact, if the electric field of the pump pulse is represented by the $\delta$ function, $p_{vc}^{0(3)}(\bm{k},\omega)$ completely cancels out $\mathcal{E}_t(\omega)$, which means that $\chi(\bm{k},\omega)$ does not have the interference term $e^{i(\omega -\Omega)\tau}$ (see Appendix A).
In contrast, when $\sigma \lesssim 1/\tau$, the coherence in the energy domain is obtained, which leads to the interference in energy space.

Second, we examine the memory effect.
When $\sigma \ll \gamma$, i.e., the pulse duration is longer than the dephasing time, $\Delta t \sim \infty$ and $\Delta E\sim 0$ are simultaneously obtained.
This leads to the relaxation that holds true as long as electrons have well-defined energies, and their energy changes are slow with the timescale of $1/\Delta \epsilon$, where $\Delta \epsilon$ is the characteristic energy exchange in a scattering event~\cite{Aihara1982, Kuznetsov1991, Kuznetsov1991_2, Rossi2002, SchaeferWegener}.
When $\sigma \gtrsim \gamma$, the relaxation involving electrons with ill-defined energies starts to contribute to the memory effect.
Therefore, if $\sigma$ and $1/\tau$ are carefully controlled to realize $1/\tau \gtrsim \sigma \gtrsim \gamma$, both coherence in the energy domain and the memory effect are relevant, and the interference in the energy domain is maintained for the time $\gamma^{-1}$.
Usually, $\gamma$ of a given system cannot be changed. However, if we make use of the quantum Zeno effect~\cite{Misra1977, Itano1990, Kaulakys1997, Streed2006}, we might be able to control $\gamma$, which can help us to observe our finding.

%%%%%%%%%%
%%%%%%%%%%
\section{Hubbard model}\label{sec_3}
Pump-probe spectroscopy has been performed in strongly correlated systems to investigate exotic phenomena~\cite{Okamoto2010, Okamoto2011, Filippis2012, Matsueda2012, Zala2013, Golez2014, Eckstein2014, Novelli2014, Prelovsek2015, Giannetti2016, Bittner2017, Bittner2018, Miyamoto2018}.
Even in correlated electron systems, there is a continuum structure in the excitation spectrum. This indicates that interference effects similar to those in the two-band model may be realized, which will be demonstrated by using a 1D half-filled Hubbard model, which is given by
\begin{align}
H=-t_\mathrm{h}\sum_{i,\sigma} \left( c^\dagger_{i,\sigma} c_{i+1,\sigma} + \mathrm{H.c.}\right) + U\sum_i n_{i,\uparrow}n_{i,\downarrow},
\label{H}
\end{align}
where $c^\dagger_{i\sigma}$ is the creation operator of an electron with spin $\sigma$ at site $i$, $n_{i,\sigma}=c^\dagger_{i,\sigma}c_{i,\sigma}$, $n_i=\sum_\sigma n_{i,\sigma}$, and $t_\mathrm{h}$ and $U$ are the nearest-neighbor hopping and on-site Coulomb interaction, respectively. Taking $t_\mathrm{h}$ to be the unit of energy ($t_\mathrm{h}=1$), we use $U=10$.

We investigate the probe-energy dependence of the optical conductivity of a Hubbard open chain with $L=10$, where $L$ is the number of sites.
We assume that both the pulses have the same shape of the vector potential given by $A(t)=A_0 e^{-(t-t_0)^2/(2t_\mathrm{d}^2)} \cos[\Omega(t-t_0)]$.
We set $A_0=0.1$, $t_0=3.0$, $t_\mathrm{d}=0.5$, and $\Omega=E_g =7.1$ for the pump pulse and $A_0=0.001$, $t_0=\tau+3.0$, $t_\mathrm{d}=0.02$, and $\Omega=E_g =7.1$ for the probe pulse unless otherwise specified, where $E_g$ is the energy of the Mott gap.
An external spatially homogeneous electric field applied along the chain in the Hamiltonian can be incorporated via the Peierls substitution in the hopping terms as $c_{i,\sigma}^\dag c_{i+1,\sigma} \rightarrow e^{iA(t)}c_{i,\sigma}^\dag c_{i+1,\sigma}$.
Using the method discussed in Refs.~\cite{Lu2015, Shao2016}, we obtain the optical conductivity in the nonequilibrium system, $\sigma(\omega,\tau) = \frac{j_\text{probe}(\omega,\tau) }{i(\omega +i\eta)LA_\text{probe}(\omega)}$, where $j_\text{probe}(\omega,\tau)$ is the Fourier transform of the current induced by the probe pulse and $A_{\text{probe}}(\omega)$ is the Fourier transform of the vector potential of the probe pulse (see Appendix B for details).

To trace the temporal evolution of the system, we employ the time-dependent Lanczos method to evaluate $|\psi (t)\rangle$. 
Here $|\psi(t+\delta{t})\rangle\simeq\sum_{l=1}^{M}{e^{-i\epsilon_l\delta{t}}}|\phi_l\rangle\langle\phi_l|\psi(t)\rangle$, where $\epsilon_l$ and $|\phi_l\rangle$ are eigenvalues and eigenvectors of the tridiagonal matrix generated in the Lanczos iteration, respectively, $M$ is the dimension of the Lanczos basis, and $\delta t$ is the minimum time step. We set $M=50$ and $\delta t=0.02$.

\begin{figure}[t]
  \centering
    \includegraphics[clip, width=20pc]{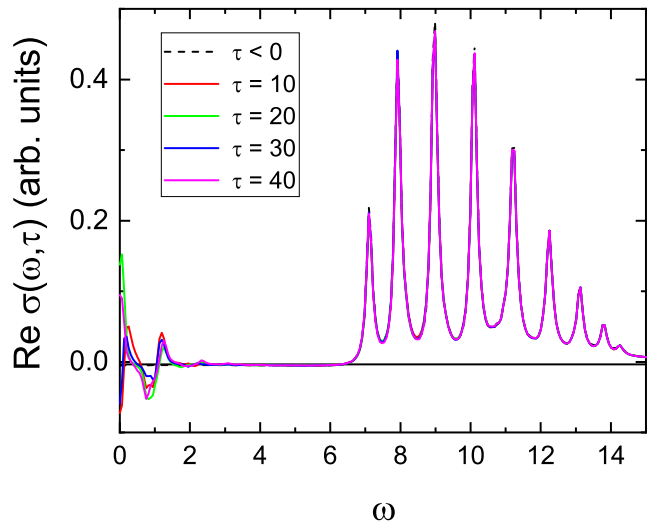}
    \caption{$\text{Re} \sigma (\omega,\tau)$ in the 1D half-filled Hubbard chain with $L=10$ and $U=10$, before pumping ($\tau<0$) and after pumping ($\tau=10$,\;20,\;30, and 40). Since the system is weakly excited, the dashed line for $\tau<0$ is almost overlapped by the solid lines above $\omega =7$.}
    \label{band}
\end{figure}

Figure~\ref{band} shows the real part of the time-dependent optical conductivity $\text{Re} \sigma(\omega,\tau)$ of the Hubbard model.
Photoinduced decreases in the spectral weights at absorption peaks above the Mott gap are small since the system is weakly excited. 
The pump photon excites carriers into an optically allowed odd-parity state.
The probe pulse couples in part to the odd-parity state, resulting in an excitation from the optically allowed state to an optically forbidden even-parity state.
In 1D Mott insulators with open boundary conditions, the optically forbidden state is located slightly above the optically allowed state~\cite{Mizuno2000}.
Low-energy in-gap excitation comes from the excitation from the optically allowed to forbidden state~\cite{Lu2015}.
Inside the Mott gap, we find photoinduced low-energy spectral weights at $\omega \simeq1.2$, 2.2, and 3.3.
These energies correspond to the energy differences between the optically allowed populated state at $\omega=7.1$ and the optically forbidden states.

\begin{figure}[t]
  \centering
    \includegraphics[clip, width=20pc]{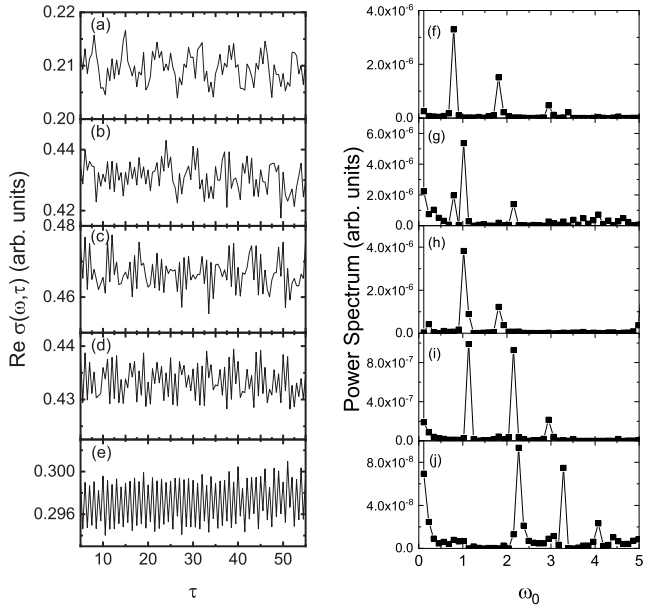}
    \caption{$\text{Re} \sigma(\omega,\tau)$ in the 1D half-filled Hubbard chain with $L=10$ and $U=10$ for (a) $\omega=7.1$, (b) $\omega=7.92$, (c) $\omega=8.98$, (d) $\omega=10.08$, and (e) $\omega=11.18$. The power spectra of $\text{Re} \sigma (\omega,\tau)$ for (f) $\omega=7.1$, (g) $\omega=7.92$, (h) $\omega=8.98$, (i) $\omega=10.08$, and (j) $\omega=11.18$.}
    \label{probe_energy_dep_w0}
\end{figure}

Figures~\ref{probe_energy_dep_w0}(a)-\ref{probe_energy_dep_w0}(e) show the $\tau$ dependence of $\text{Re} \sigma(\omega,\tau)$ above the Mott gap with probe energy $\omega=7.10$, $7.92$, $8.98$, $10.08$, and $11.18$, respectively, whose energies agree with the peak energies of the absorption spectrum in Fig.~\ref{band}.
We find that the frequencies of the oscillations depend on $\omega$.
The larger $\omega$ is, the larger the frequency is, which is consistent with our argument in the two-band model discussed above.

In order to further examine the probe-energy dependence, we show the power spectra of $\text{Re} \sigma(\omega,\tau)$ with respect to $\tau$ in Figs.~\ref{probe_energy_dep_w0}(f)-\ref{probe_energy_dep_w0}(j) for $\omega=7.1$, 7.92, 8.98, 10.08, and 11.18, respectively.
We discuss two possible contributions to the power spectra.
The first one is the contribution from the Rabi oscillation, whose frequencies are related to the low-energy in-gap states at $\omega=$1.2, 2.2, and 3.3.
In fact, we find the Rabi-oscillation contributions to the spectral weights at $\omega_0=$1.2, 2.2, and 3.3 in Figs.~\ref{probe_energy_dep_w0}(f)-\ref{probe_energy_dep_w0}(j).
Since our system is of finite size, energy levels are discretized.
Therefore, there are oscillations with resonant frequencies between the levels.
In the thermodynamic limit, the number of the levels is infinite, and thus we expect that the contributions from a huge number of such resonances with various frequencies cancel out, giving rise to an inifinite number of infinitesimal weights in the power spectra.
Thus, we consider that the Rabi-oscillation contribution to the power spectra is visible only in finite-size systems and negligible in the thermodynamic limit.

The second one is the contribution from the interference effect, which gives rise to the $\omega$ dependence of the pump-probe spectra as discussed in the two-band model. 
The oscillations with the frequencies $\omega-\Omega$ appear at $\omega_0=7.92-7.10=0.82$, $8.98-7.10=1.88$, $10.08-7.10=2.98$, and $11.18-7.10=4.08$.
These energies correspond to the energy difference between the levels at $\omega=\Omega=7.1$ and the excited states above the Mott gap, all of which belong to the same electronic states with odd parity.
We consider that this origin makes the dominant contribution to the power spectra in the thermodynamic limit.
In order to induce the transient interference, we should use the pump pulse whose spectrum covers some energy levels. Then we can store the information of the pump pulse in electronic states with a wide range of energies above the Mott gap.

According to the two possible contributions to the power spectra, in Fig.~\ref{probe_energy_dep_w0}(g), for example, we find peak structures at $\omega_0=0.82$, 1.2, and 2.2.
The peak structures at $\omega_0=1.2$ and 2.2 come from the Rabi oscillation of the two odd- and even-parity states.
On the other hand, the origin of the structure at $\omega_0=0.82$ is the interference because $\omega_0=0.82$ corresponds to one of the energy differences between the odd-odd states mentioned above.
Similarly, Figs.~\ref{probe_energy_dep_w0}(h)-\ref{probe_energy_dep_w0}(j) are understood in the same way (see Appendix B for details).

\section{electron-boson coupling in the two-band model} \label{sec_4}

Finally, we discuss the effect of bosons coupled to electrons on the probe-energy-dependent oscillation.
Nonequilibrium electron dynamics coupled to a boson driven by a laser has been extensively studied.
Furthermore, since non-Markovian relaxation is important in electron systems coupled to a bosonic environment, open quantum systems with non-Markovian properties have been studied for a long time~\cite{Jaynes1963, Caldeira1985, Leggett1987, BreuerPetruccione, Zurek2003, Reiter2014, Seetharam2015, Nazir2016, deVega2017}.
The additional Hamiltonian due to boson degrees of freedom is 
\begin{align}\label{H_ph}
\mathcal{H}_{\text{ph}} = \sum_{\bm{q}} \omega_{\bm{q}} a_{\bm{q}}^\dag a_{\bm{q}} + \sum_{\bm{k},\bm{q}} g_{\bm{q}} (a_{-\bm{q}}^\dag + a_{\bm{q}}) (c_{\text{c}\bm{k+q}}^\dag c_{\text{c}\bm{k}} + c_{\text{v}\bm{k+q}}^\dag c_{\text{v}\bm{q}}),
\end{align}
where $a_{\bm{q}}$ is an annihilation operator for bosons with momentum $\bm{q}$, $\omega_{\bm{q}}$ is the boson frequency, and $g_{\bm{q}}$ is an electron-boson coupling constant.

We examine the two-band model with electron-boson coupling under the application of the exponential pump pulse.
Total polarization is given by $p_{\text{vc}}^{}(\bm{k},t)= p_{\text{vc}}^{0}(\bm{k},t) + p_{\text{vc}}^{b}(\bm{k},t)$, where $p_{\text{vc}}^{0}(\bm{k},t)$ is from the one-particle contribution, as discussed above, and $p_{\text{vc}}^{b}(\bm{k},t)$ is from the electron-boson coupling.
Solving the kinetic equation with $\mathcal{H}_{\text{ph}}$ (see Appendix A), the probe susceptibility is given by
\begin{align} \label{Eq_chi_b_main}
&\chi ^{b}(\bm{k},\omega)\simeq \frac{p_{\text{vc}}^{b(3)}(\bm{k},\omega)}{\mathcal{E}_{\text{t}}(\omega)}\nonumber =\sum _{\bm{q}} g_{\bm{q}}^2\mathcal{N}_{\bm{q}} \cdot 4i\sigma d_{\text{cv}}\left| d_{\text{cv}}\right|{}^2 \left| \mathcal{\tilde{E}}_p\right| {}^2 \nonumber \\
&\times \Biggl[
\frac{e^{-\tau \left(\sigma-\gamma\right)} e^{i \tau  \left( -\Omega +\varepsilon_{\bm{k}} -\varrho _{\bm{k}}\right)} \left(-i\gamma -2i \Gamma - \omega + \varepsilon _{\bm{k}}- \varrho _{\bm{k}}\right) }{\left(i \gamma +\omega -\varepsilon _{\bm{k}}+\varrho _{\bm{k}}\right){}^2 \left(i \gamma +\omega -\varepsilon _{\bm{k+q}}+\varrho _{\bm{k}}+\omega _{\bm{q}}\right) v_{\bm{k}}^+}\nonumber \\
&\times \frac{\left(2 i \gamma +2 \omega -\varepsilon _{\bm{k}}-\varepsilon _{\bm{k+q}}+\varrho _{\bm{k}}+\varrho _{\bm{k-q}}+2 \omega _{\bm{q}}\right)}{(i\Gamma +i\sigma + \omega -\Omega ) \left(i \gamma +\omega -\varepsilon _{\bm{k}}+\varrho _{\bm{k-q}}+\omega _{\bm{q}}\right) u_{\bm{k}}^+u_{\bm{k}}^-}\Biggr]\nonumber \\
&+\cdots,
\end{align}
where $\mathcal{N}_{\bm{q}}=\frac{1}{e^{\omega_{\bm{q}}/k_BT}-1}$.
In the limit $\gamma \rightarrow 0$, the pole of the energy denominator $\omega = \varepsilon_{\bm{k}} - \varrho _{\bm{k}}$ gives rise to an oscillatory behavior of $e^{i(\omega -\Omega)\tau}$ with decay $e^{-(\sigma -\gamma)\tau}$, which is the same behavior as the third term in Eq.~(\ref{chi0}).
%Therefore, bosons coupled to electrons contribute to the oscillation.
Therefore, the information of pump and probe pulses is transmitted with the help of boson-assisted electron scattering, which gives one of the possible origins of the transient interference.

In Mott insulators, magnons are strongly coupled to photo-excited electrons in 2D Mott insulators, in contrast to the 1D Mott insulator where spin and charge degrees of freedom are separated.
Therefore, the interference proposed in this work will be easily realized in the 2D Mott insulators.
We thus speculate that the oscillations observed by the pump-probe spectroscopy of the 2D Mott insulator Nd$_2$CuO$_4$~\cite{Miyamoto2018} come from the interference effect. 
In order to confirm this speculation, we need to investigate theoretically the pump-probe spectrum of the 2D half-filled Hubbard model, but it remains for a future work.

\section{summary}\label{sec_5}
In summary, we suggested the transient interference in the energy domain between temporary well-separated light pulses using electronic states of band and Mott insulators as a medium, which manifests as the oscillation of the pump-probe spectrum whose frequency is indicated by Eq.~(\ref{probe_dep}).
This interference could be observed only after recent developments of ultrafast spectroscopic techniques.
The transient interference reflects the universal property of interference between two photon pulses mediated by electron systems, which does not depend on the details of the electron systems. Therefore, the interference is also realized in the presence of electron correlation since there is a continuum structure. We examined this by calculating the pump-probe spectrum in the 1D half-filled Hubbard model.
To verify our prediction, we suggested an experiment for Nd$_2$CuO$_4$ with varying pump-pulse duration and delay. 
Since our theory predicts the transient oscillation even in the 1D Mott insulators, we proposed a pump-probe experiment in Sr$_2$CuO$_3$.
Furthermore, we found that bosons coupled to electrons in the two-band model make the additional contribution to the transient interference.
Based on the result, both magnons coupled to electrons and the continuum structure in electronic excitation spectrum would be possible origins of the oscillation observed in Nd$_2$CuO$_4$.

\begin{acknowledgments}
We would like to thank H. Okamoto, T. Miyamoto, I. Eremin, and P. Prelov\v{s}ek for fruitful discussions. This work was supported by CREST (Grant No. JPMJCR1661), the Japan Science and Technology Agency, the creation of new functional devices and high-performance materials to support next-generation industries (CDMSI), and the challenge of basic science exploring extremes through multi-physics and multi-scale simulations to be tackled by using a post-K computer.
\end{acknowledgments}

%%%%%%%%%%%%%%%%%%%%%%%%%%%%%%%%%%%%
%   Appendix
%%%%%%%%%%%%%%%%%%%%%%%%%%%%%%%%%%%%
\appendix

\section{Pump-probe absorption spectrum of the two-band model}

We provide the solution of the optical Bloch equations (\ref{Eq_bloch_eq_1}) and (\ref{Eq_bloch_eq_2}) and derive the pump-probe absorption spectrum.
With the assumption of dipole transitions, the Hamiltonian (\ref{H}) of the two-band model under the time-dependent electric field reads
\begin{align}
\mathcal{H} =& \sum_{\bm{k}} \varepsilon_{\bm{k}} c_{\text{c}\bm{k}}^\dag c_{\text{c}\bm{k}} + \sum_{\bm{k}} \varrho_{\bm{k}} c_{\text{v}\bm{k}}^\dag c_{\text{v}\bm{k}} \nonumber \\
&-\sum_{\bm{k}} \left( d_{\text{cv}} \mathcal{E}(t) c_{\text{c}\bm{k}}^\dag c_{\text{v}\bm{k}} + d_{\text{cv}}^* \mathcal{E}(t) c_{\text{v}\bm{k}}^\dag c_{\text{c}\bm{k}} \right),
\end{align}
where $c_{\text{c(v)}\bm{k}}$ is an annihilation operator for fermions in the conduction (valence) band with momentum $\bm{k}$. The energies of the conduction and valence bands are $\varepsilon_{\bm{k}} = \varepsilon + \frac{\hbar ^2 \bm{k}^2}{2m_{\text{c}}},\; \varrho_{\bm{k}} = \varrho + \frac{\hbar ^2 \bm{k}^2}{2m_{\text{v}}},$
where $\varepsilon$ and $\varrho$ are the minimum and maximum of the conduction and valence band, respectively, and $m_c$ and $m_v$ are the effective masses of electrons in the conduction and valence bands, respectively.
We introduce the interband dipole matrix element $d_{\text{cv}}$ and external electric field $\mathcal{E}(t)$. Hereafter, we set $\hbar=1$.
Taking the long-wave-length limit of electric field, the optical Bloch equations (\ref{Eq_bloch_eq_1}) and (\ref{Eq_bloch_eq_2}) are written as
\begin{align}
&\left( \frac{\partial}{\partial t} + i\{ \varepsilon _{\bm{k}}-\varrho_{\bm{k}}-i \gamma \} \right) p_{\text{vc}}^0(\bm{k},t) = d_{\text{cv}} \mathcal{E}(t) \{ 1-2f_c(\bm{k}) \} 
\end{align}
and
\begin{align}
\left(  \frac{\partial}{\partial t} +  \Gamma \right) f_{\text{c}}(\bm{k},t) = -2\text{Im}\left[ d_{\text{cv}} \mathcal{E}(t) p_{\text{vc}}^{0*}(\bm{k},t) \right],
\end{align}
where $f_{\text{c}}(\bm{k}) = \langle c_{\text{c}\bm{k}}^\dag c_{\text{c}\bm{k}} \rangle$ and $p_{\text{vc}}^0(\bm{k}) = \langle c_{\text{v}\bm{k}}^\dag c_{\text{c}\bm{k}} \rangle$, where $\langle \cdots \rangle$ represents the expectation value.
We introduce a phenomenological damping rate $\Gamma$ for $f_{\text{c}}$, and a dephasing rate $\gamma$ for $p_{\text{vc}}^0$.
We consider an electric field $\mathcal{E}(t) = \frac{1}{2}\left(\mathcal{\tilde{E}}(t)e^{-i\Omega t} + \mathcal{\tilde{E}}^*(t)e^{i\Omega t}\right)$, where $\mathcal{\tilde{E}}(t)=2\left\{ \mathcal{\tilde{E}}_{\text{p}}(t)e^{i\bm{k}_p\cdot \bm{r}} + \mathcal{\tilde{E}}_{\text{t}}(t)e^{i\bm{k}_t\cdot \bm{r}} \right\}$, and the electric field and wave vector of the pump (probe) pulse are $\mathcal{\tilde{E}}_{\text{p}}$ and $\bm{k}_{\text{p}}$ ($\mathcal{\tilde{E}}_{\text{t}}$ and $\bm{k}_{\text{t}}$).
Introducing an expansion parameter $\lambda$ through $\mathcal{E}(t) \rightarrow \lambda \mathcal{E}(t)$, we obtain $p_{\text{vc}}^0 = \lambda p_{\text{vc}}^{0(1)} + \lambda^2 p_{\text{vc}}^{0(2)} + \lambda^3 p_{\text{vc}}^{0(3)} + \cdots,\; f_{\text{c}} = \lambda f_{\text{c}}^{(1)} + \lambda^2 f_{\text{c}}^{(2)} + \lambda^3 f_{\text{c}}^{(3)} + \cdots$. 
The shape of the probe pulse is represented by the delta function, $\mathcal{\tilde{E}}_{\text{t}}(t) = \mathcal{\tilde{E}}_{\text{t}} \delta (t -\tau)$ $(\tau >0)$, where $\tau$ is delay time between the pump and probe pulses.
With the rotating-wave approximation, $\tilde{p}_{\text{vc}}^{0(3)}(\bm{k},t)= \tilde{p}_{\text{vc,A}}^{0(3)}(\bm{k},t)+\tilde{p}_{\text{vc,B}}^{0(3)}(\bm{k},t)$ is given by
\begin{widetext} 
\begin{align}
\label{p0A}
\tilde{p}_{\text{vc,A}}^{0(3)} (\bm{k},t)
=& -i2d_{\text{cv}} |d_{\text{cv}}|^2 e^{i\bm{k}_{t}\cdot \bm{r}} \int _{-\infty} ^t dt' e^{-i\left\{ \varepsilon_{\bm{k}}-\varrho_{\bm{k}}-\Omega -i\gamma \right\}(t-t')} \mathcal{\tilde{E}}_{p}(t') e^{-\Gamma (t'-\tau)}\mathcal{\tilde{E}}_{t} \theta (t'-\tau) \nonumber \\
&\times \int_{-\infty} ^{\tau} dt''' e^{i\left\{ \varepsilon_{\bm{k}}-\varrho_{\bm{k}}-\Omega -i\gamma \right\}(\tau-t''')}\mathcal{\tilde{E}}_{p}^*(t''') \nonumber \\
&-i2d_{\text{cv}} |d_{\text{cv}}|^2 e^{i\bm{k}_{t}\cdot \bm{r}} \int _{-\infty} ^t dt' e^{-i\left\{ \varepsilon_{\bm{k}}-\varrho_{\bm{k}}-\Omega -i\gamma \right\}(t-t')} \mathcal{\tilde{E}}_{p}(t') \int _{-\infty} ^{t'} dt'' e^{-\Gamma (t'-t'')} \mathcal{\tilde{E}}_{p}^*(t'') \nonumber \\
&\times \mathcal{\tilde{E}}_{t} \theta (t''-\tau) e^{-i\left\{ \varepsilon_{\bm{k}}-\varrho_{\bm{k}}-\Omega -i\gamma \right\}(t''-\tau)}
\end{align}
and
\begin{align}
\label{p0B}
\tilde{p}_{\text{vc,B}}^{0(3)}(\bm{k},t) %=& -\int _{-\infty}^t dt' e^{-i\left\{ \varepsilon_{\bm{k}}-\varrho_{\bm{k}}-\Omega -i\gamma \right\}(t-t')} i2d_{\text{cv}} \mathcal{\tilde{E}}_{t}(t')e^{i\bm{k}_{t}\cdot \bm{r}}f_{\text{c}}^{(2)}(\bm{k},t')\nonumber \\
=& -i2d_{\text{cv}} |d_{\text{cv}}|^2 e^{i\bm{k}_{t}\cdot \bm{r}} e^{-i\left\{ \varepsilon_{\bm{k}}-\varrho_{\bm{k}}-\Omega -i\gamma \right\}(t-\tau)} \mathcal{\tilde{E}}_{t} \theta (t-\tau) \int _{-\infty} ^{\tau} dt'' \mathcal{\tilde{E}}_{p}(t'') e^{-\Gamma (\tau-t'')} \nonumber \\
&\times \int_{-\infty} ^{t''} dt''' e^{i\left\{ \varepsilon_{\bm{k}}-\varrho_{\bm{k}}-\Omega -i\gamma \right\}(t''-t''')}\mathcal{\tilde{E}}_{p}^*(t''') \nonumber \\
&-i2d_{\text{cv}} |d_{\text{cv}}|^2 e^{i\bm{k}_{t}\cdot \bm{r}} e^{-i\left\{ \varepsilon_{\bm{k}}-\varrho_{\bm{k}}-\Omega -i\gamma \right\}(t-\tau)} \mathcal{\tilde{E}}_{t} \theta (t-\tau) \int _{-\infty} ^{\tau} dt'' \mathcal{\tilde{E}}_{p}^*(t'') e^{-\Gamma (\tau-t'')} \nonumber \\
&\times \int _{-\infty} ^{t''} dt'''  \mathcal{\tilde{E}}_{p}(t'') e^{-i\left\{ \varepsilon_{\bm{k}}-\varrho_{\bm{k}}-\Omega -i\gamma \right\}(t''-t''')},
\end{align}
\end{widetext}
where we are interested in contributions with a phase factor $e^{i\bm{k}_{t}\cdot \bm{r}}$, i.e., in the direction of the probe beam.
We include only terms which are linear in $\mathcal{\tilde{E}}_t$, and ignore all terms that are higher than second order in $\mathcal{\tilde{E}}_p$.
We use the delta function $\mathcal{\tilde{E}}_{t}(t) = \mathcal{\tilde{E}}_{t} \delta (t-\tau)$ to represent a probe pulse. 
%At times shorter than the dephasing time, the carriers are not in definite-energy eigenstates and therefore cannot be described classically.

Taking $\mathcal{\tilde{E}}_{p}(t) = \mathcal{\tilde{E}}_{p} \delta (t)$, from Eqs.~(\ref{p0A}) and (\ref{p0B}), we obtain
\begin{align}
&\tilde{p}_{\text{vc}}^{0(3)}(\bm{k},t)=\tilde{p}_{\text{vc,B}}^{0(3)}(\bm{k},t)\nonumber \\
=&-2 i d_{\text{cv}} \theta(\tau) \mathcal{\tilde{E}}_t \left| d_{\text{cv}} \right| {}^2 \left| \mathcal{\tilde{E}}_p \right| {}^2 \theta(t-\tau ) e^{-\Gamma \tau +i \bm{k}_t\cdot \bm{r}+\left(-\gamma -i \Delta_{\bm{k}} \right) (t-\tau )},
\end{align}
where $\Delta_{\bm{k}} = \varepsilon _{\bm{k}} - \varrho _{\bm{k}} -\Omega$.
The Fourier transformation of $p_{\text{vc}}^{0(3)}(\bm{k},t)$ is given by
\begin{align}
&p_{\text{vc}}^{0(3)}(\bm{k},\omega)=\int _{-\infty} ^{\infty} dt e^{i\omega t} p_{\text{vc}}^{0(3)}(\bm{k},t) \nonumber \\
&= \frac{2 d_ {\text{cv}} \theta (\tau ) \mathcal{\tilde{E}}_t \left| d_{\text{cv}} \mathcal{\tilde{E}}_p \right | {}^2 e^{i \left[\bm{k}_t\cdot \bm{r}+\ \tau (i \Gamma +\omega -\Omega )\right]}}{i \gamma-\varepsilon_{\bm{k}}+\varrho_{\bm{k}}+\omega }.
\end{align}
The probe susceptibility reads
\begin{align}
\chi(\bm{k},\omega)\simeq \frac{p_{\text{vc}}^{0(3)}(\bm{k},\omega)}{\mathcal{E}_t(\omega)}=\frac{2 d_{\text{cv}} \theta (\tau )  \left| d_{\text{cv}} \mathcal{\tilde{E}}_p \right| {}^2 e^{-\Gamma  \tau }}{i \gamma-\varepsilon_{\bm{k}}+\varrho_{\bm{k}}+\omega },
\end{align}
where $\mathcal{E}_t(\omega)=\int _{-\infty} ^\infty dt\mathcal{E}_t(t)e^{i\omega t} \simeq \mathcal{\tilde{E}}_t e^{i(\omega -\Omega) \tau} e^{i\bm{k}_t\cdot \bm{r}}$.
Since the oscillatory term $e^{i(\omega -\Omega)\tau}$ of $p_{vc}^{0(3)}(\bm{k},\omega)$ cancels out that of the probe electric field $\mathcal{E}_t(\omega)$, $\chi(\bm{k},\omega)$ does not have terms depending on $e^{i(\omega -\Omega) \tau}$.
%Here, we have to note that in the case of $\tau<0$, namely pump pulse follows probe pulse, the situation is changed and .
%This is related to the so-called ``coherent artifact"

However, if we consider a pump pulse written by $\mathcal{\tilde{E}}_p(t)=\mathcal{\tilde{E}}_p e^{-\sigma|t|}$, we obtain
\begin{widetext}
\begin{align}
&\tilde p_{\text{vc}}^{0(3)}(\bm{k},t) \nonumber \\
=&2 i d_ {\text{cv}} \mathcal{\tilde{E}}_t \left| d_{\text{cv}}\right| {}^2 \left| \mathcal{\tilde{E}}_p \right| {}^2 \theta (t-\tau ) e^{i \bm{k}_t\cdot \bm{r}} \nonumber\\
&\times \Biggl[
   \frac{i e^{-(\gamma +i \Delta_{\bm{k}} ) (t-\tau )} \left(-\frac{e^{-2
   \sigma  t}-e^{-2 \sigma  \tau }}{2 \sigma }+\frac{e^{-2 \sigma  \tau }-e^{t (\gamma
   -\Gamma +i \Delta_{\bm{k}} -\sigma )-\tau  (\gamma -\Gamma +i \Delta_{\bm{k}} +\sigma )}}{\gamma -\Gamma
   +i \Delta_{\bm{k}} -\sigma }\right)}{-i \gamma +i \Gamma +\Delta_{\bm{k}} }\nonumber \\
   &-\frac{\left(\sigma 
   \left(-1+2 e^{\tau  (\gamma +i \Delta_{\bm{k}} +\sigma )}\right)+\gamma +i \Delta_{\bm{k}} \right)
   e^{\tau  (\Gamma -\sigma )-t (\gamma +i \Delta_{\bm{k}} )} \left(e^{t (\gamma -\Gamma +i \Delta_{\bm{k}}
   -\sigma )}-e^{\tau  (\gamma -\Gamma +i \Delta_{\bm{k}} -\sigma )}\right)}{(\gamma +i \Delta_{\bm{k}}
   -\sigma ) (\gamma +i \Delta_{\bm{k}} +\sigma ) (\gamma -\Gamma +i \Delta_{\bm{k}} -\sigma
   )}\nonumber \\
   &+\left(-\frac{e^{\tau  (\Gamma -2 \sigma )}-1}{(\Gamma -2 \sigma ) (\gamma +i \Delta_{\bm{k}}
   +\sigma )}-\frac{2 \sigma  \left(-1+e^{\tau  (\gamma +\Gamma +i \Delta_{\bm{k}} -\sigma
   )}\right)}{(\gamma +i \Delta_{\bm{k}} -\sigma ) (\gamma +i \Delta_{\bm{k}} +\sigma ) (\gamma +\Gamma +i
   \Delta_{\bm{k}} -\sigma )}-\frac{1}{(\Gamma +2 \sigma ) (\gamma +i \Delta_{\bm{k}} -\sigma )}\right)\nonumber \\
&\times e^{-\Gamma  \tau -(\gamma +i \Delta_{\bm{k}} ) (t-\tau )}\nonumber \\
   &+\left(-\frac{e^{\tau  (\Gamma -2
   \sigma )}-1}{(\Gamma -2 \sigma ) (-\gamma -i \Delta_{\bm{k}} +\sigma )}-\frac{2 \sigma  e^{-\tau
    (\gamma -\Gamma +i \Delta_{\bm{k}} +\sigma )} \left(-1+e^{\tau  (\gamma -\Gamma +i \Delta_{\bm{k}}
   +\sigma )}\right)}{(\gamma +i \Delta_{\bm{k}} -\sigma ) (\gamma +i \Delta_{\bm{k}} +\sigma ) (\gamma
   -\Gamma +i \Delta_{\bm{k}} +\sigma )}+\frac{1}{(\Gamma +2 \sigma ) (\gamma +i \Delta_{\bm{k}} +\sigma
   )}\right)\nonumber \\
&\times e^{-\Gamma  \tau -(\gamma +i \Delta_{\bm{k}} ) (t-\tau )}
\Biggr].
\end{align}
\end{widetext}
\begin{widetext}
The probe susceptibility is given by
\begin{align}
&\chi (\bm{k},\omega)\simeq\frac{p_{\text{vc}}^{0(3)}(\bm{k},\omega)}{\mathcal{E}_t(\omega)}\nonumber \\
=&
\frac{id_{\text{cv}}\left|d_{\text{cv}}\right| {}^2 \left| \mathcal{\tilde{E}}_p\right| {}^2 }{u_{\bm{k}}^+ u_{\bm{k}}^-} \Biggl[\frac{4 e^{- (\gamma +\sigma ) \tau} e^{i \tau  \left(\Omega - \varepsilon _{\bm{k}}+\varrho   _{\bm{k}}\right)} \sigma }{\left(i \gamma +\omega -\varepsilon _{\bm{k}}+\varrho _{\bm{k}}\right) v_{\bm{k}}^- }
-\frac{4 e^{-\left(\sigma - \gamma \right)\tau} e^{i \tau  \left( - \Omega +\varepsilon _{\bm{k}}-\varrho   _{\bm{k}}\right)} \sigma }{(i\Gamma +i\sigma + \omega -\Omega ) v_{\bm{k}}^+ } 
-\frac{8i e^{-\left(\sigma - \gamma \right)\tau} e^{i \tau  \left( - \Omega +\varepsilon _{\bm{k}}-\varrho   _{\bm{k}}\right)}  \Gamma \sigma }{\left(i \gamma +\omega -\varepsilon _{\bm{k}}+\varrho _{\bm{k}}\right)(i\Gamma +i\sigma + \omega -\Omega ) v_{\bm{k}}^+ }\nonumber \\
&+e^{- \Gamma \tau}(\cdots) +e^{-2 \sigma \tau} \left(\cdots\right)\Biggr],
\end{align}
\end{widetext}
where $u_{\bm{k}}^{\pm}=i\gamma \pm i\sigma + \Omega - \varepsilon _{\bm{k}} + \varrho _{\bm{k}}$, $v_{\bm{k}}^{\pm} = i\gamma \pm i\Gamma \mp i\sigma + \Omega - \varepsilon _{\bm{k}}+ \varrho _{\bm{k}}$, and $(\cdots)$ represents an abbreviation of the $\tau$-independent part of the corresponding term.
The third term is shown in Eq.~(\ref{chi0}).
%Therefore, if a pump pulse has finite width, the oscillatory term with $e^{i(\omega -\Omega)\tau}$ remains, which corresponds to the experimental observation.

Next, we consider the contribution from electrons coupled to bosons to the interference.
The additional Hamiltonian (\ref{H_ph}) due to boson degrees of freedom is 
\begin{align*}
\mathcal{H}_{\text{ph}} = \sum_{\bm{q}} \omega_{\bm{q}} a_{\bm{q}}^\dag a_{\bm{q}} + \sum_{\bm{k},\bm{q}} g_{\bm{q}} (a_{-\bm{q}}^\dag + a_{\bm{q}}) (c_{\text{c}\bm{k+q}}^\dag c_{\text{c}\bm{k}} + c_{\text{v}\bm{k+q}}^\dag c_{\text{v}\bm{q}}),
\end{align*}
where $a_{\bm{q}}$ is an annihilation operator for bosons with momentum $\bm{q}$, $\omega_{\bm{q}}$ is the boson frequency, and $g_{\bm{q}}$ is an electron-boson coupling constant.
Total polarization is given by $p_{\text{vc}}^{}(\bm{k},t)= p_{\text{vc}}^{0}(\bm{k},t) + p_{\text{vc}}^{b}(\bm{k},t)$, where $p_{\text{vc}}^{0}(\bm{k},t)$ is from the one-particle contribution, as discussed above, and $p_{\text{vc}}^{b}(\bm{k},t)$ is from the electron-boson coupling.

If carriers are created by optical pulses, the wave function is a superposition of states in the conduction and valence bands.
As long as this phase coherence is maintained, i.e., at times shorter than the dephasing time, the carriers are not in definite-energy eigenstates, which requires the non-Markovian description of relaxation.
To obtain the quantum kinetic equation with non-Markovian relaxation, we use the Keldysh nonequilibrium Green's function that is a two-time generalization of the density matrix.
Two characteristic timescales of the scattering time and the duration of the interaction process determine the dynamics of the carriers.
The optical Bloch equation with electron-boson coupling is given by using the nonequilibrium Green's function and reads
\begin{align}
&\left( \frac{\partial}{\partial t} + i\{ \varepsilon _{\bm{k}}-\varrho_{\bm{k}}-i \gamma \} \right) p_{\text{vc}}(\bm{k},t) \nonumber \\
=& d_{\text{cv}} \mathcal{E}(\bm{r},t) \{ 1-2f_{\text{c}}(\bm{k}) \} \nonumber \\
&+(-i) \sum _{\bm{q}} \left[ g_{\bm{q}}^2\mathcal{N}_{\bm{q}} \{ P_{\text{vc}} ^+ (\bm{k},\bm{k+q},t) - P_{\text{vc}}^+(\bm{k-q},\bm{k},t) \} \right] \nonumber \\
&+ (-i) \sum _{\bm{q}} [\mathcal{N}_{\bm{q}} \leftrightarrow \mathcal{N}_{\bm{q}}+1, \omega_{\bm{q}} \leftrightarrow -\omega_{\bm{q}}],
\end{align}
\begin{align}
&\left( \frac{\partial}{\partial t}+i\{ \varepsilon_{\bm{k+q}} -\varrho_{\bm{k}} - \omega _{\bm{q}} -i \gamma\}\right)P_{\text{vc}}^+(\bm{k},\bm{k+q},t) \nonumber \\
=& i\{p_{\text{vc}}(\bm{k+q},t) - p_{\text{vc}}(\bm{k},t)\},
\end{align}
and
\begin{align}
\left(  \frac{\partial}{\partial t} +  \Gamma \right) f_{\text{c}}(\bm{k},t) =& -2\text{Im}\left[ d_{\text{cv}} \mathcal{E}(t) p_{\text{vc}}^*(\bm{k},t) \right],
\end{align}
where $P^+_{\text{vc}}(\bm{k},\bm{k+q},t)$ are boson-assisted excitonic transitions, $\gamma$ accounts for all dephasing processes other than electron-boson scattering, and $\mathcal{N}_{\bm{q}}=\frac{1}{e^{\omega_{\bm{q}}/k_BT}-1}$ is a thermal magnon distribution~\cite{HaugJauho,SchaeferWegener}.
Solving the equation of motion, we obtain
\begin{align}
\label{pb}
&p_{\text{vc}}^b(\bm{k},t)\nonumber \\
=& \int _{-\infty}^t dt' e^{-i\left\{\varepsilon _{\bm{k}}-\varrho_{\bm{k}}-i \gamma  \right\}(t-t')} \nonumber \\
&\times (-i) \Biggl[ \sum _{\bm{q}} \left[ g_{\bm{q}}^2\mathcal{N}_{\bm{q}} \{ P_{\text{vc}} ^+ (\bm{k},\bm{k+q},t') - P_{\text{vc}}^+(\bm{k-q},\bm{k},t') \} \right]\nonumber \\
&+ \sum _{\bm{q}} [\mathcal{N}_{\bm{q}} \leftrightarrow \mathcal{N}_{\bm{q}}+1, \omega_{\bm{q}} \leftrightarrow -\omega_{\bm{q}}] \Biggr],
\end{align}
where the last term means the replacement of $\mathcal{N}_{\bm{q}}$ with $\mathcal{N}_{\bm{q}}+1$ and $\omega_{\bm{q}}$ with $-\omega_{\bm{q}}$ in the previous terms.
$P_{\text{vc}}^{+(3)}(\bm{k},\bm{k+q},t)$ and $p_{\text{vc}}^{b(3)}(\bm{k},t)$ are written as
\begin{align}
P_{\text{vc}}^{+(3)}(\bm{k},\bm{k+q},t) =& \int _{-\infty}^t du e^{-i (t-u) \left(\varepsilon _ {\bm{k+q}}-i \gamma-\omega_{\bm{q}}-\varrho _{\bm{k}}\right)}\nonumber \\
&\times i[p_{\text{vc}}^{(3)}(\bm{k+q},u)-p_{\text{vc}}^{(3)}(\bm{k},u)]\nonumber\\
\simeq& \int _{-\infty}^t du e^{-i (t-u) \left(\varepsilon _ {\bm{k+q}}-i \gamma-\omega_{\bm{q}}-\varrho _{\bm{k}}\right)}\nonumber \\
&\times i[p_{\text{vc}}^{0 (3)}(\bm{k+q},u)-p_{\text{vc}}^{0 (3)}(\bm{k},u)]\nonumber \\
\end{align}
and
\begin{align}
p_{\text{vc}}^{b(3)}(\bm{k},t)=&(-i)\sum _{\bm{q}} g_{\bm{q}}^2\mathcal{N}_{\bm{q}} \int _{-\infty} ^t du e^{-i (\varepsilon _{\bm{k}} -\varrho _{\bm{k}} -i \gamma ) (t-u) }\nonumber \\
&\times \left\{P_{\text{vc}}^{+(3)}(\bm{k},\bm{k+q},u) - P_{\text{vc}}^{+(3)} (\bm{k-q},\bm{k},u)\right\}\nonumber \\
&+[\mathcal{N}_{\bm{q}}\leftrightarrow \mathcal{N}_{\bm{q}}+1,\omega_{\bm{q}} \leftrightarrow -\omega_{\bm{q}}],\nonumber \\
\end{align}
respectively.
If $\mathcal{\tilde{E}}_p(t)=\mathcal{\tilde{E}}_p\delta(t)$ is used, we obtain
\begin{align}
&\chi^{b}(\bm{k},\omega)\simeq\frac{p_{\text{vc}}^{b(3)}(\bm{k},\omega)}{\mathcal{E}_t(\omega)}\nonumber \\
=&\sum _{\bm{q}} g_{\bm{q}}^2\mathcal{N}_{\bm{q}} \frac{2 d_{\text{cv}} \theta (\tau )  \left| d_{\text{cv}}\right| {}^2 \left| \mathcal{\tilde{E}}_p\right| {}^2 e^{-\Gamma \tau}}{\left(i \gamma -\varepsilon _{\bm{k}}+\varrho _{\bm{k}}+\omega
   \right){}^2}\nonumber \\
&\times \Biggl[\frac{\varepsilon
   _{\bm{k}}-\varepsilon _{\bm{k-q}}+\varrho _{\bm{k-q}}-\varrho _{\bm{k}}}{\left(i \gamma -\varepsilon _{\bm{k-q}}+\varrho
   _{\bm{k-q}}+\omega \right) \left(i \gamma +\omega _{\bm{q}}-\varepsilon _{\bm{k}}+\varrho _{\bm{k-q}}+\omega
   \right)}\nonumber \\
&-\frac{-\varepsilon _{\bm{k}}+\varepsilon _{\bm{k+q}}-\varrho _{\bm{k+q}}+\varrho _{\bm{k}}}{\left(i \gamma
   -\varepsilon _{\bm{k+q}}+\varrho _{\bm{k+q}}+\omega \right) \left(i \gamma +\omega _{\bm{q}}-\varepsilon
   _{\bm{k+q}}+\varrho _{\bm{k}}+\omega \right)} \Biggr]\nonumber \\
&+[\mathcal{N}_{\bm{q}}\leftrightarrow \mathcal{N}_{\bm{q}}+1,\omega_{\bm{q}} \leftrightarrow -\omega_{\bm{q}}].
\end{align}
When the pump pulse is represented by the $\delta$ function, we cannot obtain the oscillating term $e^{i(\omega -\Omega)\tau}$, even if we take into account the boson-assisted transition.

If the pump pulse is written by $ \mathcal{E}_p(t)=\mathcal{E}_p e^{-\sigma|t|}$, we obtain
\begin{widetext}
\begin{align}
&\chi ^{b}(\bm{k},\omega)\simeq \frac{p_{\text{vc}}^{b(3)}(\bm{k},\omega)}{\mathcal{E}_t(\omega)}\nonumber \\
=&\sum _{\bm{q}} g_{\bm{q}}^2\mathcal{N}_{\bm{q}} \cdot 4i\sigma d_{\text{cv}}\left| d_{\text{cv}}\right|{}^2  \left| \mathcal{\tilde{E}}_p\right| {}^2 \nonumber \\
&\times \Biggl[
\frac{e^{-\tau \left(\sigma-\gamma\right)} e^{i \tau  \left( -\Omega +\varepsilon_{\bm{k}} -\varrho _{\bm{k}}\right)} \left(-i\gamma -2i \Gamma - \omega + \varepsilon _{\bm{k}}- \varrho _{\bm{k}}\right) \left(2 i \gamma +2 \omega -\varepsilon _{\bm{k}}-\varepsilon _{\bm{k+q}}+\varrho _{\bm{k}}+\varrho _{\bm{k-q}}+2 \omega _{\bm{q}}\right) }{(i\Gamma +i\sigma + \omega -\Omega ) \left(i \gamma +\omega -\varepsilon _{\bm{k}}+\varrho _{\bm{k}}\right){}^2 \left(i \gamma +\omega -\varepsilon _{\bm{k+q}}+\varrho _{\bm{k}}+\omega _{\bm{q}}\right) \left(i \gamma +\omega -\varepsilon _{\bm{k}}+\varrho _{\bm{k-q}}+\omega _{\bm{q}}\right)v_{\bm{k}}^+ u_{\bm{k}}^+ u_{\bm{k}}^-}\nonumber \\
&+\frac{e^{-\left(\sigma-\gamma\right)\tau} e^{i \tau \left(-\Omega+\varepsilon _{\bm{k+q}} -\varrho _{\bm{k+q}}\right)} \left(i\gamma +2i \Gamma + \omega - \varepsilon _{\bm{k+q}}+ \varrho _{\bm{k+q}}\right) }{(i\Gamma + i\sigma + \omega -\Omega ) \left(i \gamma +\omega -\varepsilon _{\bm{k}}+\varrho _{\bm{k}}\right)  \left(i \gamma +\omega -\varepsilon _{\bm{k+q}}+\varrho _{\bm{k+q}}\right) \left(i \gamma +\omega -\varepsilon _{\bm{k+q}}+\varrho _{\bm{k}}+\omega _{\bm{q}}\right) v_{\bm{k+q}}^+ u_{\bm{k+q}}^+ u_{\bm{k+q}}^-}\nonumber \\
&+\frac{e^{- \tau \left(\sigma-\gamma\right)} e^{i \tau \left( - \Omega+\varepsilon _{\bm{k-q}} -\varrho _{\bm{k-q}}\right)} \left(i\gamma +2i \Gamma + \omega - \varepsilon _{\bm{k-q}}+ \varrho _{\bm{k-q}}\right)}{(i\Gamma +i\sigma + \omega -\Omega ) \left(i \gamma +\omega -\varepsilon _{\bm{k}}+\varrho _{\bm{k}}\right) \left(i \gamma +\omega -\varepsilon _{\bm{k-q}}+\varrho _{\bm{k-q}}\right) \left(i \gamma +\omega -\varepsilon _{\bm{k}}+\varrho _{\bm{k-q}}+\omega _{\bm{q}}\right) v_{\bm{k-q}}^+ u_{\bm{k-q}}^+ u_{\bm{k-q}}^-}\nonumber \\
&+\frac{e^{- (\gamma+\sigma)\tau } e^{i \tau  \left(\Omega-\varepsilon _{\bm{k}} +\varrho _{\bm{k}}\right)}\left(2 i \gamma +2 \omega -\varepsilon _{\bm{k}}-\varepsilon _{\bm{k+q}}+\varrho _{\bm{k}}+\varrho _{\bm{k-q}}+2 \omega _{\bm{q}}\right) }{\left(-i\gamma - \omega + \varepsilon _{\bm{k}}- \varrho _{\bm{k}}\right){}^2 \left(-i\gamma - \omega + \varepsilon_{\bm{k}}- \varrho _{\bm{k-q}}- \omega _{\bm{q}}\right) \left(-i\gamma -\omega + \varepsilon _{\bm{k+q}}- \varrho _{\bm{k}}- \omega _{\bm{q}}\right) v_{\bm{k}}^- u_{\bm{k}}^+ u_{\bm{k}}^-}\nonumber \\
&-\frac{e^{- \left(\gamma +\sigma\right)\tau} e^{i \tau  \left(\Omega-\varepsilon _{\bm{k+q}} +\varrho _{\bm{k+q}}\right)}}{\left(-i\gamma - \omega + \varepsilon _{\bm{k}}- \varrho _{\bm{k}}\right) \left(-i\gamma - \omega + \varepsilon _{\bm{k+q}}- \varrho _{\bm{k+q}}\right) \left(-i\gamma - \omega + \varepsilon _{\bm{k+q}}- \varrho _{\bm{k}}- \omega _{\bm{q}}\right) v_{\bm{k+q}}^- u_{\bm{k+q}}^+ u_{\bm{k+q}}^-}\nonumber \\
&+\frac{e^{-\left( \gamma +\sigma \right)\tau} e^{i \tau \left(\Omega-\varepsilon _{\bm{k-q}} +\varrho _{\bm{k-q}}\right)}}{\left(-i\gamma - \omega + \varepsilon _{\bm{k}}- \varrho _{\bm{k}}\right) \left(-i\gamma - \omega + \varepsilon _{\bm{k-q}}- \varrho _{\bm{k-q}}\right) \left(-i\gamma - \omega + \varepsilon _{\bm{k}}- \varrho _{\bm{k-q}}- \omega_{\bm{q}}\right) v_{\bm{k-q}}^- u_{\bm{k-q}}^+ u_{\bm{k-q}}^-}\Biggr]\nonumber \\
&+[\mathcal{N}_{\bm{q}}\leftrightarrow \mathcal{N}_{\bm{q}}+1,\omega_{\bm{q}} \leftrightarrow -\omega_{\bm{q}}]+(\cdots).
\end{align}
\end{widetext}
The first term is shown in Eq.~(\ref{Eq_chi_b_main}).

\section{Time-dependent optical conductivity in the nonequilibrium state}

Using the method discussed in Refs~\cite{Lu2015, Shao2016}, we obtain optical conductivities in the nonequilibrium system. 
In order to identify the response of the system with respect to later probe pulses, subtraction is necessary; that is, two successive steps are involved in order to calculate the optical conductivity in nonequilibrium. First, a time-evolution process that describes the nonequilibrium development of the system in the absence of a probe pulse is evaluated, which gives rise to $ j_\text{pump}(t)$.
Second, in the presence of an additional probe pulse, we get $j_\text{total}(t,\tau)$. 
The subtraction of $j_\text{pump}(t)$ from $j_\text{total}(t,\tau)$ produces the required $j_\text{probe}(t,\tau)$, i.e., the variation of the current expectations due to the presence of the probe pulse.
Then, the optical conductivity in nonequilibrium is proposed to be
\begin{align}
\sigma(\omega,\tau) = \frac{j_\text{probe}(\omega,\tau) }{i(\omega +i\eta)LA_\text{probe}(\omega)},
\end{align}
where $A_{\text{probe}}(\omega)$ is the Fourier transform of the vector potential of probe pulses and $L$ is the number of sites.

We find photoinduced spectral weights at $\omega \simeq1.2$, 2.2, and 3.3 inside the Mott gap as shown in Fig.~\ref{band}.
Low-energy in-gap excitation comes from the excitation from the optically allowed to forbidden state~\cite{Lu2015}.
These energies correspond to the energy differences between the optically allowed populated state with the odd parity at $\omega=7.1$ and the optically forbidden states with the even parity.

Figures~2(h)-2(j) can be understood as discussed in Sec.~\ref{sec_3}. 
In Fig.~2(h), we find peak structures at $\omega_0=1.2$ and 1.88.
The frequency $\omega_0=1.2$ corresponds to the energy of a photoinduced low-energy state, which comes from the Rabi oscillation of the odd- and even-parity states.
The origin of the structure at $\omega_0=1.88$ comes from the interference effect that gives rise to the energy difference between the two states with the odd parity, i.e., $\omega-\Omega=8.98-7.10=1.88$.
In Fig.~2(i), we find peak structures at $\omega_0=1.2$, 2.2, and 2.98.
The frequencies $\omega_0=1.2$ and 2.2 correspond to the energy of photoinduced low-energy states.
The origin of the structure at $\omega_0=2.98$ comes from the interference effect.
In Fig.~2(j), we find peak structures at $\omega_0=2.2$, 3.3, and 4.08.
The frequencies $\omega_0=2.2$ and 3.3 correspond to the energy of photoinduced low-energy states.
The origin of the structure at $\omega_0=4.08$ is the interference effect.

If we take $\Omega=7.92$, which is larger than the Mott gap of $7.10$, we also find the peak structures that come from the interference effect.
We find that the oscillations with the frequencies $|\omega-\Omega|$ appear at $\omega_0=|7.10-7.92|=0.82$, $8.98-7.92=1.06$, $10.08-7.92=2.16$, and $11.18-7.92=3.26$.
Therefore, the oscillation due to the transient interference appears even when $\omega-\Omega < 0$.
This property may give useful information for experiments on the transient interference in the pump-probe spectroscopy of the Mott insulators~\cite{Kawakami}.

In order to obtain the contribution from the interference, three conditions are needed.
First, we should not impose a step function on the vector potential, but rather the Gaussian function to give the electric field of the probe pulse.
Although the same optical conductivity is obtained in equilibrium by using the two kinds of vector potentials of the probe pulse, this is not the case in nonequilibrium~\cite{Shao2016}.
If we impose the step function on the vector potential of the probe pulse, we cannot obtain the spectral weights originating from the interference.
An oscillating probe field with a central frequency will be needed to interfere with a pump pulse.
Second, in order to generate the interference, the frequencies of the pump and probe pulses should be (nearly) the same.
Third, the spectral width of the pump pulse should not be too small.
The cooperation of electronic states in the band structure is important for maintaining the information of the pump pulse.
To excite electronic states with a wide range of energy above the Mott gap, we have to use the pump pulse whose spectrum covers several energy levels.


\begin{thebibliography}{99} 
\bibitem{Mukamel1995} S. Mukamel, \textit{Principles of Nonlinear Optical Spectroscopy} (Oxford University Press, New York, 1995).
\bibitem{Diels1996} J.-C. Diels and W. Rudolph, \textit{Ultrashort Laser Phenomena} (Academic, New York, 1996).
\bibitem{Krausz2009} F. Krausz and M. Ivanov, \href{https://doi.org/10.1103/RevModPhys.81.163}{Rev. Mod. Phys. \textbf{81}, 163 (2009).} % Attosecond physics
\bibitem{Giannetti2016} C. Giannetti, M. Capone, D. Fausti, M. Fabrizio, F. Parmigiani, and D. Mihailovic, \href{https://doi.org/10.1080/00018732.2016.1194044}{Adv. Phys. \textbf{65}, 58 (2016).}
\bibitem{Wollenhaupt2002} M. Wollenhaupt, A. Assion, D. Liese, Ch. Sarpe-Tudoran, T. Baumert, S. Zamith, M. A. Bouchene, B. Girard, A. Flettner, U. Weichmann, and G. Gerber, \href{https://doi.org/10.1103/PhysRevLett.89.173001}{Phys. Rev. Lett. \textbf{89}, 173001 (2002).} % Interferences of Ultrashort Free Electron Wave Packets
\bibitem{Lindner2005} F. Lindner, M. G. Sch\"{a}tzel, H. Walther, A. Baltu\v{s}ka, E. Goulielmakis, F. Krausz, D. B. Milo\v{s}evi\'{c}, D. Bauer, W. Becker, and G. G. Paulus, \href{https://doi.org/10.1103/PhysRevLett.95.040401}{Phys. Rev. Lett. \textbf{95}, 040401 (2005).} % Attosecond Double-Slit Experiment
\bibitem{Kiffner2006} M. Kiffner, J. Evers, and C. H. Keitel, \href{https://doi.org/10.1103/PhysRevLett.96.100403}{Phys. Rev. Lett. \textbf{96}, 100403 (2006).} % Quantum Interference Enforced by Time-Energy Complementarity
\bibitem{Milosevic2006} D. B. Milo\v{s}evi\'{c}, G. G. Paulus, D. Bauer, and W. Becker, \href{https://doi.org/10.1088/0953-4075/39/14/R01}{J. Phys. B \textbf{39} R203 (2006).} % Above-threshold ionization by few-cycle pulses
\bibitem{Kruger2018} M. Kr\"{u}ger, C. Lemell , G. Wachter, J. Burgd\"{o}rfer, and P. Hommelhoff, \href{https://doi.org/10.1088/1361- 6455/aac6ac}{J. Phys. B \textbf{51} 172001 (2018).} % Attosecond physics phenomena at nanometric tips
\bibitem{Fleischauer2002} M. Fleischhauer and M. D. Lukin, \href{https://doi.org/10.1103/PhysRevA.65.022314}{Phys. Rev. A \textbf{65}, 022314 (2002).} % Quantum memory for photons: Dark-state polaritons
\bibitem{Carlson1983} N. W. Carlson, L. J. Rothberg, A. G. Yodh, W. R. Babbitt, and
T. W. Mossberg, \href{https://doi.org/10.1364/OL.8.000483}{Opt. Lett. \textbf{8}, 483 (1983).} % Storage and time reversal of light pulses using photon echoes
\bibitem{Leung1982} K. P. Leung, T. W. Mossberg, and S. R. Hartmann, \href{https://doi.org/10.1016/0030-4018(82)90110-9}{Opt. Commun. \textbf{43}, 145 (1982).} % OBSERVATION AND DENSITY DEPENDENCE OF THE RAMAN ECHO IN ATOMIC THALLIUM VAPOR 
\bibitem{Hemmer1994} P. R. Hemmer, K. Z. Cheng, J. Kierstead, M. S. Shariar, and M. K. Kim, \href{https://doi.org/10.1364/OL.19.000296}{Opt. Lett. \textbf{19}, 296 (1994).} % Time-domain optical data storage by use of Raman coherent population trapping
\bibitem{Ohmori2009} K. Ohmori, \href{https://doi.org/10.1146/annurev.physchem.59.032607.093818}{Annu. Rev. Phys. Chem. \textbf{60}, 487 (2009).} % Wave-Packet and Coherent Control Dynamics

\bibitem{Miyamoto2018} T. Miyamoto, Y. Matsui, T. Terashige, H. Yada, S. Ishihara, Y. Watanabe, S. Adachi, T. Ito, K. Oka, A. Sawa, and H. Okamoto, \href{https://doi.org/10.1038/s41467-018-06312-z}{Nat. Commun. \textbf{9}, 3948 (2018).}
\bibitem{HaugKoch} H. Haug and S. Koch, \textit{Quantum Theory of the Optical and Electronic Properties of Semiconductors} (World Scientific, Singapore, 2004).
\bibitem{Rhodes2013} M. Rhodes, G. Steinmeyer, J. Ratner, and R. Trebino, \href{https://doi.org/10.1002/lpor.201200102}{Laser Photonics Rev. \textbf{7}, 557 (2013).} % Coherent artifact
\bibitem{Englert1996} B.-G. Englert, \href{https://doi.org/10.1103/PhysRevLett.77.2154}{Phys. Rev. Lett. \textbf{77}, 2154 (1996).} % Fringe Visibility and Which-Way Information: An Inequality
\bibitem{Durr1998} S. D\"{u}rr, T. Nonn, and G. Rempe, \href{https://doi.org/10.1038/25653}{Nature (London) \textbf{395}, 33 (1998).} % Origin of quantum-mechanical complementarity probed by a `which-way' experiment in an atom interferometer
\bibitem{SchaeferWegener} W. Sch\"{a}fer and M. Wegener, \textit{Semiconductor Optics and Transport Phenomena} (Springer, Berlin, 2002).
\bibitem{Kuznetsov1991} A. V. Kuznetsov, \href{https://doi.org/10.1103/PhysRevB.44.13381}{Phys. Rev. B \textbf{44}, 13381 (1991).} % Coherent and non-Markovian effects in ultrafast relaxation of photoexcited hot carriers: A model study
\bibitem{Kuznetsov1991_2} A. V. Kuznetsov, \href{https://doi.org/10.1103/PhysRevB.44.8721}{Phys. Rev. B \textbf{44}, 8721 (1991).}
\bibitem{Aihara1982} M. Aihara, \href{https://doi.org/10.1103/PhysRevB.25.53}{Phys. Rev. B \textbf{25}, 53 (1982).} % Non-Markovian theory of nonlinear-opticai phenomena associated with the extremely fast relaxation in condensed matter
\bibitem{Rossi2002} F. Rossi and T. Kuhn, \href{https://doi.org/10.1103/RevModPhys.74.895}{Rev. Mod. Phys. \textbf{74}, 895 (2002).} % Theory of ultrafast phenomena in photoexcited semiconductors
\bibitem{Misra1977} B. Misra and E. C. G. Sudarshan, \href{https://doi.org/10.1063/1.523304}{J. Math. Phys. Sci. \textbf{18}, 756 (1977).}
\bibitem{Itano1990} W. M. Itano, D. J. Heinzen, J. J. Bollinger, and D. J. Wineland, \href{https://doi.org/10.1103/PhysRevA.41.2295}{Phys. Rev. A \textbf{41}, 2295 (1990).}
\bibitem{Kaulakys1997} B. Kaulakys and V. Gontis, \href{https://doi.org/10.1103/PhysRevA.56.1131}{Phys. Rev. A \textbf{56}, 1131 (1997).}
\bibitem{Streed2006} E. W. Streed, J. Mun, M. Boyd, G. K. Campbell, P. Medley, W. Ketterle, and D. E. Pritchard, \href{https://doi.org/10.1103/Phys- RevLett.97.260402}{Phys. Rev. Lett. \textbf{97}, 260402 (2006).}
% Mott insulator
\bibitem{Okamoto2010} H. Okamoto, T. Miyagoe, K. Kobayashi, H. Uemura, H. Nishioka, H. Matsuzaki, A. Sawa, and Y. Tokura, \href{https://doi.org/10.1103/PhysRevB.82.060513}{Phys. Rev. B \textbf{82}, 060513 (2010).} % Ultrafast charge dynamics in photoexcited Nd2CuO4 and La2CuO4 cuprate compounds investigated by femtosecond absorption spectroscopy
\bibitem{Okamoto2011} H. Okamoto, T. Miyagoe, K. Kobayashi, H. Uemura, H. Nishioka, H. Matsuzaki, A. Sawa, and Y. Tokura, \href{https://doi.org/10.1103/PhysRevB.83.125102}{Phys. Rev. B \textbf{83}, 125102 (2011).} % Photoinduced transition fromMott insulator to metal in the undoped cuprates Nd2CuO4 and La2CuO4
\bibitem{Filippis2012}G. De Filippis, V. Cataudella, E. A. Nowadnick, T. P. Devereaux, A. S. Mishchenko, and N. Nagaosa, \href{https://doi.org/10.1103/PhysRevLett.109.176402}{Phys. Rev. Lett. {\bf 109}, 176402 (2012).}
\bibitem{Matsueda2012} H. Matsueda, S. Sota, T. Tohyama, and S. Maekawa, \href{https://doi.org/10.1143/JPSJ.81.013701}{J. Phys. Soc. Jpn. \textbf{81}, 013701 (2012).}
\bibitem{Zala2013} Z. Lenar\v{c}i\v{c} and P. Prelov\v{s}ek, \href{https://doi.org/10.1103/PhysRevLett.111.016401}{Phys. Rev. Lett. \textbf{111}, 016401 (2013).} % Ultrafast Charge Recombination in a Photoexcited Mott-Hubbard Insulator 
\bibitem{Golez2014} D. Gole\v{z}, J. Bon\v{c}a, M. Mierzejewski, and L. Vidmar, \href{https://doi.org/10.1103/PhysRevB.89.165118}{Phys. Rev. B \textbf{89}, 165118 (2014).} % Mechanism of ultrafast relaxation of a photo-carrier in antiferromagnetic spin background
\bibitem{Eckstein2014} M. Eckstein and P. Werner, \href{https://doi.org/10.1103/PhysRevLett.113.076405}{Phys. Rev. Lett. \textbf{113}, 076405 (2014).} % Ultrafast Separation of Photodoped Carriers in Mott Antiferromagnets
\bibitem{Novelli2014} F. Novelli, G. De Filippis, V. Cataudella, M. Esposito, I. Vergara, F. Cilento, E. Sindici, A. Amaricci, C. Giannetti, D. Prabhakaran, S. Wall, A. Perucchi, S. Dal Conte, G. Cerullo, M. Capone, A. Mishchenko, M. Gr\"uninger, N. Nagaosa, F. Parmigiani, and D. Fausti, \href{https://doi.org/10.1038/ncomms6112}{Nat. Commun. \textbf{5}, 5112 (2014).}
\bibitem{Prelovsek2015} P. Prelov\v{s}ek, J. Kokalj, Z. Lenar\v{c}i\v{c}, and R. H. McKenzie, \href{https://doi.org/10.1103/PhysRevB.92.235155}{Phys. Rev. B \textbf{92}, 235155 (2015).} % Holon-doublon binding as the mechanism for the Mott transition
\bibitem{Bittner2017} N. Bittner, T. Tohyama, S. Kaiser, and D. Manske, \href{https://arxiv.org/abs/1706.09366}{arXiv:1706.09366} % Light–induced superconductivity in a strongly correlated electron system
\bibitem{Bittner2018} N. Bittner, D. Gole\v{z}, H. U. R. Strand, M. Eckstein, and P. Werner, \href{https://doi.org/10.1103/PhysRevB.97.235125}{Phys. Rev. B 97, 235125 (2018).} % Coupled charge and spin dynamics in a photo-excitedMott insulator
\bibitem{Lu2015} H. Lu, C. Shao, J. Bon\v{c}a, D. Manske, and T. Tohyama, \href{https://doi.org/10.1103/PhysRevB.91.245117}{Phys. Rev. B \textbf{91}, 245117 (2015).}
\bibitem{Shao2016} C. Shao, T. Tohyama, H.-G. Luo, and H. Lu, \href{https://doi.org/10.1103/PhysRevB.93.195144}{Phys. Rev. B \textbf{93}, 195144 (2016).}
\bibitem{Mizuno2000} Y. Mizuno, K. Tsutsui, T. Tohyama, and S. Maekawa, \href{https://doi.org/10.1103/PhysRevB.62.R4769}{Phys. Rev. B \textbf{62}, R4769(R) (2000).} % Nonlinear optical response and spin-charge separation in one-dimensional Mott insulators
\bibitem{Jaynes1963} E. T. Jaynes and F. W. Cummings, \href{https://doi.org/10.1109/PROC.1963.1664}{Proc. IEEE \textbf{51}, 89 (1963).} % Comparison of quantum and semiclassical radiation theories with application to the beam maser
\bibitem{Caldeira1985} A. O. Caldeira and A. J. Leggett, \href{https://doi.org/10.1103/PhysRevA.31.1059}{Phys. Rev. A \textbf{31}, 1059 (1985).} % Influence of damping on quantum interference: An exactly soluble model
\bibitem{Leggett1987} A. J. Leggett, S. Chakravarty, A. T. Dorsey, M. P. A. Fisher, A. Garg, and W. Zwerger, \href{https://doi.org/10.1103/RevModPhys.59.1}{Rev. Mod. Phys. \textbf{59}, 1 (1987).} % Dynamics of the dissipative two-state system
\bibitem{BreuerPetruccione} H.-P. Breuer and F. Petruccione, \textit{The Theory of Open Quantum Systems} (Oxford University Press, Oxford, 2002).
\bibitem{Zurek2003} W. H. Zurek, \href{https://doi.org/10.1103/RevModPhys.75.715}{Rev. Mod. Phys. \textbf{75}, 715 (2003).} % Decoherence, einselection, and the quantum origins of the classical
\bibitem{Reiter2014} D. E. Reiter, T. Kuhn, M. Gl\"{a}ssl, and V. M. Axt, \href{https://doi.org/10.1088/0953-8984/26/42/423203}{J. Phys.: Condens. Matter \textbf{26}, 423203 (2014).} % The role of phonons for exciton and biexciton generation in an optically driven quantum dot
\bibitem{Seetharam2015} K. I. Seetharam, C.-E. Bardyn, N. H. Lindner, M. S. Rudner, and G. Refael, \href{https://doi.org/10.1103/PhysRevX.5.041050}{Phys. Rev. X \textbf{5}, 041050 (2015).} % Controlled Population of Floquet-Bloch States via Coupling to Bose and Fermi Baths
\bibitem{Nazir2016} A. Nazir and D. P. S. McCutcheon, \href{https://doi.org/10.1088/0953-8984/28/10/103002}{J. Phys.: Condens. Matter \textbf{28} 103002 (2016).} % Modelling exciton-phonon interactions in optically driven quantum dots
\bibitem{deVega2017} I. de Vega and D. Alonso, \href{https://doi.org/10.1103/RevModPhys.89.015001}{Rev. Mod. Phys. \textbf{89}, 015001 (2017).} % Dynamics of non-Markovian open quantum systems
\bibitem{HaugJauho} H. Haug and A.-P. Jauho, \textit{Quantum Kinetics in Transport and Optics of Semiconductors} (Springer, Berlin, 1996).
\bibitem{Kawakami} Y. Kawakami (private communication).
\end{thebibliography}
\end{document}